# Advanced Thermostats for Molecular Dynamics

Roumen Tsekov

Department of Physical Chemistry, University of Sofa, 1164 Sofia, Bulgaria

Advanced thermostats for molecular dynamics are proposed on the base of the rigorous Langevin dynamics. Because the latter accounts for the subsystem-bath interactions in detail, the bath anisotropy and nonuniformity are described via the relevant friction tensor. The developed model reflects properly the relativistic dynamics of the subsystem evolution as well as the nonlinear friction, which can occur for fast particles with large momenta at elevated temperature.

According to quantum mechanics the most complete description of a quantum system is given in terms of the wave function. Accordingly, the phase space probability density $\rho$ from classical mechanics is replaced by the density matrix operator $\hat{\rho}$, which obeys the von Neumann equation. In the frames of the Schrödinger picture its formal solution in the energy basis acquires the form $\hat{\rho} = \sum\sum \exp[i(E_j - E_k)t/\hbar]|k\rangle\langle k|\hat{\rho}_0|j\rangle\langle j|$, where $\{E_k\}$ is the set of the energy eigenvalues of the system Hamiltonian. As is seen the density matrix possesses nondiagonal elements, in contrast to the equilibrium density operator $\hat{\rho}_{eq} = \sum p_k |k\rangle\langle k|$, which is diagonal. According to quantum statistical physics $p_k$ is the probability for occupation of the state with energy $E_k$. Clearly, the density operator $\hat{\rho}$ is recurrent function of time, which reflects the Poincare cycles. In fact, evolution never stops, and the stationary equilibrium distribution is an idealization, when fluctuations are somehow omitted. It is evident that one could eliminate the effect of the persistent fluctuations by averaging over time and thus the equilibrium distribution is the most frequently occupied one. Integrating the density matrix operator $\hat{\rho}$ on time leads straightforward to $\hat{\rho}_{eq} = \sum |k\rangle\langle k|\hat{\rho}_0|k\rangle\langle k|$, where non-degenerated energy spectrum of the system is assumed. Identifying the initial probability density $p_k = \langle k|\hat{\rho}_0|k\rangle = \delta_{E_k E}$ from the system energy conservation, the density operator reduces to the expression $\hat{\rho}_{eq} = \delta_{\hat{H}E}$, known from the equilibrium quantum statistical physics for the quantum Gibbs microcanonical ensemble. This result shows that the decoherence in isolated systems is caused by the quantum evolution itself and the averaging in time leads to mutual cancelation of the non-diagonal fluctuating elements. It is expected that the self-decoherence mechanism takes place in open quantum systems as well, thus assisting the decoherence caused by the environment [1].

In classical mechanics the Newtonian dynamics also preserves energy and according to the Poincare recurrence theorem, a closed mechanical system returns arbitrarily close to its initial state. Such quasiperiodic systems are conveniently described via the action-angle coordinates [2]. Because the Hamilton function of quasiperiodic systems depends solely on the action $I$, the integration of the relevant Hamilton equations $\dot{\varphi} = \partial_I H$ and $\dot{I} = -\partial_\varphi H$ is simple. The action remains constant equal to the initial value $I_0$, while the angle $\varphi = \varphi_0 + \omega_0 t$ increases linearly in time with frequency $\omega_0 \equiv \partial_{I_0} H$. The microscopic probability density acquires the following form $\rho = \delta(I - I_0)\delta(\varphi - \varphi_0 - \omega_0 t)$, which is a continuous function of time. Hence, $\rho$ is fluctuating permanently, which reflects the lack of equilibrium distribution in the time-reversible mechanics. Like quantum mechanics the equilibrium thermodynamic state could be attributed to the most frequently observed microscopic state. Therefore, the equilibrium thermodynamic distribution is the time-averaged microscopic probability density

$$\rho_{eq} \equiv \lim_{\tau \to \infty} \int_0^\tau \rho(I, \varphi, t) dt / \tau = \delta(I - I_0) \lim_{\tau \to \infty} \int_0^\tau \delta(\varphi - \varphi_0 - \omega_0 t) dt / \tau \qquad (1)$$

Employing the properties of the Dirac delta-function, Eq. (1) simplifies to $\rho_{eq} = \delta(\omega_0 \cdot (I - I_0))/\Omega$ where $\Omega$ reflects the average periods of the dynamic oscillations. Introducing the system energy via $E \equiv H(I_0)$, the equilibrium distribution acquires the form $\rho_{eq} = \delta(H - E)/\Omega$ of the classical Gibbs microcanonical distribution. As is seen, all nonadditive integrals of motion vanish and the system energy remains the only one involved. It follows from the derivation above that the average value of any quantity by ensemble coincides with the average value on time. Therefore, the ergodic theorem is always fulfilled for quasiperiodic systems. An interesting consequence from our analysis is that the time averaging can result in some metaphysical correlations. Imagine two noninteracting subsystems are set together. While they are statistically independent in mechanics, because $\rho(1 \cap 2) = \rho(1)\rho(2)$, a statistical correlation appears from Eq. (1) in thermodynamics, since $\rho_{eq}(1 \cap 2) \neq \rho_{eq}(1)\rho_{eq}(2)$, which could explain the positive entropy of mixing in thermodynamics. Perhaps the KAM theory can through light on statistical interference between almost noninteracting systems such as entanglement [1, 3]. According to Eq. (1) the thermodynamic equilibrium state is a superposition of the most frequently observed (most probable) mechanical states. Such a picture corresponds to the Boltzmann point of view and supports the time-coarse-grained solution of the entropy production problem. This is not surprising, because any thermodynamic measurement requires finite time, which is always larger than the system resonances. The latter are, in general, truly short in many particles systems.

If one considers now an open mechanical subsystem, a general problem is how the thermal bath affects the subsystem equilibrium distribution. The rigorous way to answer this question is to consider the subsystem $S$ and bath $B$ as closed mechanical system, which possesses naturally the Gibbs microcanonical distribution at equilibrium. Because the unified system is isolated, its Hamilton function can be written as the sum $H_{SB} = H_S + H_B + U_{SB}$, where $U_{SB}$ is the potential, describing the subsystem-bath interaction. Integrating the classical Gibbs microcanonical distribution $\rho_{eq} = \delta(H_{SB} - E)/\Omega$ along the bath particles momenta $P$ and coordinates $Q$ yields the equilibrium distribution of the subsystem particles

$$f_{eq} \equiv \int \rho_{eq} dPdQ \sim \int [1-(H_S + U_B + U_{SB})/E]^{3N_B/2} dQ \qquad (2)$$

where the last expression is derived via explicit integration over the bath particles momenta. The number of bath particles $N_B$ tends to infinity in thermodynamic limit and the energy becomes linear function of it. Thus, using the limit $\lim_{N_B \to \infty} E/N_B = 3k_B T/2$, Eq. (2) acquires the form

$$f_{eq} \sim \int \exp[-\beta(H_S + U_B + U_{SB})]dQ = \exp[-\beta(H_S + A_{BS})]/Z \qquad (3)$$

where $\beta = 1/k_B T$ is the reciprocal bath temperature and $Z$ is the subsystem partition function. One can easily recognize that $A_{BS}(q)$ is the conditional configurational Helmholtz free energy of the bath at a fixed configuration $q$ of the subsystem particles. It describes the average effect of the subsystem-bath interaction and, in contrast to the usual potentials, $A_{BS}$ could be temperature dependent. The Cooper pairs in superconductors are very popular manifestation of interactions, mediated by the bath particles. Other examples are electrostatic screening and the Friedel oscillations. Note that the generalized Gibbs canonical distribution (3) does not obey the subsystem Liouville equation, because $\{H_S, f_{eq}\} \neq 0$. Since the interactions among particles depend only on the distance between them, $A_{BS}$ is constant if the subsystem consists of a single particle. Therefore, the Maxwell-Boltzmann distribution is not affected by the subsystem-bath interaction.

The Newton equations of classical mechanics preserve energy and for this reason they are suitable for description of molecular dynamics in microcanonical ensemble. On the other hand, experimental measurements are usually conducted at constant temperature in the presence of thermostats. Hence, it is a challenge for the theory to define classical mechanics for open systems and there are several attempts proposed in the literature, such as the Andersen, Berendsen and Nose-Hoover thermostats [4-6]. The goal of the present paper is rigorously to start the analysis from the exact Langevin dynamics of complex mechanical systems. According to classical mechanics the Hamilton function $H_S(p,q)$ completely defines the behavior of a mechanical subsystem of $N$ particles in the free state without thermostat. The corresponding Hamilton equations of motion $\dot{q} = \partial_p H_S$ and $\dot{p} = -\partial_q H_S$ describe the evolution of the $3N$-dimensional vectors of momenta $p$ and coordinates $q$ of all particles. If the subsystem is part of a larger mechanical system, where the rest is referred as the thermal bath, tree additional forces appear strictly in the Langevin equation [7-9]

$$\dot{q} = \partial_p H \qquad \dot{p} = -\partial_q H - B \cdot \dot{q} + F \qquad (4)$$

The extended Hamiltonian function $H = H_S + A_{BS}$ accounts for possible thermostat mediated interactions via the bath conditional free energy $A_{BS}$, which depends on temperature. The friction tensor $B$ of the dissipative force, which is proportional to the velocity $\dot{q}$, reflects frictional anisotropy of the bath. The dependence of $B(p,q)$ on the momenta and coordinates accounts for frictional nonlinearity and nonuniformity, respectively. The friction force controls the subsystem energy loss via heat transfer to the bath. The reverse energy flux generates the fluctuation force $F(p,q,t)$, which is stochastic, because the state of the bath particles is unknown. It follows directly from Eq. (4), however, that the evolution of the subsystem probability density $f(p,q,t)$ in the phase space obeys the Klein-Kramers equation [7-9]

$$\partial_t f + \partial_p H \cdot \partial_q f - \partial_q H \cdot \partial_p f = \partial_p \cdot B \cdot (f \partial_p H + k_B T \partial_p f) \qquad (5)$$

where the constant bath temperature $T$ emerges from the fluctuation-dissipation theorem. The equilibrium solution of Eq. (5) $f_{eq} = \exp(-\beta H)/Z$ is the Gibbs canonical distribution (3), where $Z$ is the corresponding partition function and $\beta \equiv 1/k_B T$ is the reciprocal bath temperature.

At equilibrium, the right-hand side of Eq. (5) is zero, which is a result of dynamic counterbalance between two opposite rates of dissipation and fluctuation of energy, respectively. Therefore, a new dynamic quantity $\zeta$ can be introduced via

$$3N\dot{\zeta} = \partial_p \cdot (k_B T B \cdot \partial_p f_{eq})/f_{eq} = -\partial_p \cdot (f_{eq} B \cdot \partial_p H)/f_{eq} \qquad (6)$$

It follows directly from its definition that $<\dot{\zeta}>_{eq} = 0$. Employing the Gibbs canonical distribution, one can further transform Eq. (6) to the following thermo-mechanical equation

$$\dot{\zeta} = (\beta \partial_p H - \partial_p) \cdot B \cdot \partial_p H / 3N \qquad (7)$$

which involves the reciprocal bath temperature $\beta$ as well. To elucidate the physical meaning of Eq. (7), one can calculate via Eq. (5) the average energy production $<\dot{H}> = -3N k_B T <\dot{\zeta}>$, which reveals $k_B <\zeta>$ as the bath entropy changes, caused by any degree of motion of the subsystem particles.

The rigorous way for the subsystem thermalization described above is too complicated for direct numerical simulations because the state of the bath particles is unknown. A possibility to approximate the problem is to model the bath-subsystem dynamic force $F - B \cdot \dot{q} = -\zeta B \cdot \dot{q}$ as proportional to the friction force, scaled by the bath entropy fluctuations. The stronger fluctuation force $F$ is driven by entropy decrease $\zeta < 0$ in the surroundings, while the bath entropy increase is due to prevailing dissipation. Obviously, $\zeta$ governs the direction of the heating/cooling process and $<\zeta>_{eq} = 0$ guarantees the fluctuation-dissipation theorem. Thus, the exact Eq. (4) simplifies accordingly to

$$\dot{q} = \partial_p H \qquad \dot{p} = -\partial_q H - \zeta B \cdot \dot{q} \qquad (8)$$

which describe dissipative dynamics with the fluctuating friction factor $\zeta$. Since the latter evolves in accordance with Eq. (7), the system of these equations is a pure mechanical problem, which is easier to simulate via molecular dynamics. To check consistency, one can consider the extended distribution density $f(p,q,\zeta,t)$, which accounts for the entropy fluctuations as well. Following Eq. (8), the relevant evolutionary equation reads

$$\partial_t f + \partial_p H \cdot \partial_q f - \partial_q H \cdot \partial_p f = \zeta \partial_p \cdot (fB \cdot \partial_p H) - \partial_\zeta (\dot\zeta f) \qquad (9)$$

By employing Eq. (7) one can show that $f_{eq} = (3N/2\pi)^{1/2} \exp(-3N\zeta^2/2)\exp(-\beta H)/Z$ is the equilibrium solution of Eq. (9). Therefore, the developed approximation leads to the rigorous Gibbs canonical distribution, while the front part of $f_{eq}$ represents the Gaussian distribution of the equilibrium entropy fluctuations. As expected, their dispersion $<\zeta^2>_{eq} = 1/3N$ obeys the usual law from statistical thermodynamics.

The application of the developed complex model requires knowledge for the friction tensor, which is the main coupling parameter of the subsystem to the bath. The dependence of $B(p,q)$ on the momenta $p$ is more bizarre and marks violation of the linear friction regime [8]. One can mention here, for instance, the Amontons-Coulomb law with $B \sim M/|p|$, where $M$ the diagonal mass matrix of the subsystem particles. Another very general model follows from the Frenkel theory of the activated transport, where the friction force is proportional to the inverse hyperbolic sine from the particles' momenta. Thus, at large momenta the cubic friction becomes essential [10], while at slow velocity the Ohmic friction takes place. In the latter case of slow motion, the friction tensor does not depend on momenta $p$. If the media is isotropic, the friction tensor $B = M\gamma$ is diagonal. However, if the bath is structured, the collision frequency $\gamma(q)$ is modulated by the bath-subsystem interactions. Its dependence on the positions of the subsystem particles could strongly affect the dynamics of the latter. For example, if the subsystem evolves in a solid, e.g. a zeolite, $\gamma$ is a periodic function of $q$, which reflects the symmetry of the bath [9]. Because in the nonrelativistic case the subsystem Hamilton function is $H = p \cdot M^{-1} \cdot p/2 + U(q)$ where the effective potential $U$ could also be modulated by the interaction with the bath particles via $A_{BS}$ [9], Eqs. (7-8) reduce to

$$p = M \cdot \dot q \qquad \dot p = -\partial_q U - \zeta\gamma p \qquad \dot\zeta = \gamma(\beta \dot q \cdot p/3N - 1) \qquad (10)$$

It is proven by Hoover that at constant $\gamma$ Eq. (10) resembles the Nose-Hoover thermostat [11]. Hence, the model from Eqs. (7-8) represents a relativistic Nose-Hoover thermostat, accounting also for anisotropy, nonlinearity and nonuniformity of the friction between the subsystem particles and the thermal bath. Obviously, the mechanical quantity $\theta \equiv \dot{q} \cdot p / 3Nk_B$ plays the role of the subsystem temperature and $<\theta>_{eq} = T$. The differential nature of Eq. (7) essentially complicates the molecular dynamics simulations, and for this reason many researchers are looking for simpler thermostats. Assuming equality of the reduced head exchanged between the subsystem and the bath, one can write $F/T = B \cdot \dot{q}/\theta$. Thus, a simpler mechanical alternative of the stochastic Langevin equation (4) reads

$$\dot{q} = \partial_p H \qquad \dot{p} = -\partial_q H - (1 - T/\theta) B \cdot \dot{q} \qquad (11)$$

In the nonrelativistic case with linear friction Eq. (11) predicts that the subsystem energy evolves as $\dot{H} = 3N\gamma k_B(T - \theta)$. Hence, the bath entropy production $\dot{\zeta} = -\dot{H}/3Nk_B T = \gamma(\theta/T - 1)$ satisfies again Eq. (10). Obviously, Eq. (11) represents a generalized Berendsen thermostat [12], which is not reproducing, however, the Gibbs canonical distribution because $1 - T/\theta \neq \dot{\zeta}$.

[1] Tsekov, R. (2022) Quantum entanglement of free particles, Fluct. Noise Lett., **21**, 2250024.

[2] Arnold, V. I. (1989) Mathematical Methods of Classical Mechanics, Moscow, Nauka.

[3] El-Nabulsi, R. A. (2018) Nonlocal approach to nonequilibrium thermodynamics and nonlocal heat diffusion processes, Contin. Mech. Thermodyn., **30**, 889-915.

[4] Frenkel, D., B. Smit (2002) Understanding Molecular Simulation: From Algorithms to Applications, London, Academic Press.

[5] Hünenberger, P. H. (2005) Thermostat algorithms for molecular dynamics simulations, Adv. Polym. Sci., **173**, 105-149.

[6] Leimkuhler, B., C. Matthews (2015) Molecular Dynamics, Heidelberg, Springer.

[7] Coffey, W. T., Yu. P. Kalmykov (2017) The Langevin Equation: With Applications to Stochastic Problems in Physics, Chemistry and Electrical Engineering, Singapore, World Scientific.

[8] Klimontovich, Yu. L. (1994) Nonlinear Brownian motion, Phys. Usp., **37**, 737-766.

[9] Tsekov, R., E. Ruckenstein (1994) Stochastic dynamics of a subsystem interacting with a solid body with application to diffusive processes in solids, J. Chem. Phys., **100**, 1450-1455.

[10] Hoover, W. G., B. L. Holian (1996) Kinetic moments method for the canonical ensemble distribution, Phys. Lett. A, **211**, 253-257.

[11] Hoover, W. G. (1985) Canonical dynamics: Equilibrium phase-space distributions, Phys. Rev. A, **31**, 1695-1697.

[12] Berendsen, H. J. C., J. P. M. Postma, W. F. van Gunsteren, A. Di Nola, J. R. Haak (1984) Molecular dynamics with coupling to an external bath, J. Chem. Phys., **81**, 3684-3690.